\documentclass[conference]{IEEEtran}
\IEEEoverridecommandlockouts
\usepackage{cite}
\usepackage{amsmath,amssymb,amsfonts}
\usepackage{algorithmic}
\usepackage{graphicx}
\usepackage{textcomp}
\usepackage{xcolor}
\usepackage{array}
\usepackage{mdwmath}
\usepackage{mdwtab}
\usepackage{eqparbox}
\usepackage{url}
\usepackage{braket}
\usepackage{cite}
\usepackage{doi}
\usepackage{hyperref}

\def\BibTeX{{\rm B\kern-.05em{\sc i\kern-.025em b}\kern-.08em
    T\kern-.1667em\lower.7ex\hbox{E}\kern-.125emX}}

\makeatletter
\newcommand{\linebreakand}{%
  \end{@IEEEauthorhalign}
  \hfill\mbox{}\par
  \mbox{}\hfill\begin{@IEEEauthorhalign}
}
\makeatother

\begin{document}

\title{Impact of AC Magnetic Field on Decoherence of Quantum Dot based Single Spin Qubit System\\
\thanks{This work is supported by the Department of Science and Technology
(DST), India with grant no. DST/INSPIRE/04/2018/000023.}
}

\author{\IEEEauthorblockN{Tanmay Sarkar}
\IEEEauthorblockA{\textit{Department of Electrical Engineering} \\
\textit{Indian Institute of Technology, Roorkee}\\
Roorkee, India \\
tsarkar@ee.iitr.ac.in}
\and
\IEEEauthorblockN{Yash Tiwari}
\IEEEauthorblockA{\textit{Department of Electronics and Communication Engineering} \\
\textit{Indian Institute of Technology, Roorkee}\\
Roorkee, India\\
ytiwari@ec.iitr.ac.in}
\linebreakand
\IEEEauthorblockN{Vishvendra Singh Poonia}
\IEEEauthorblockA{\textit{Department of Electronics and Communication Engineering} \\
\textit{Indian Institute of Technology, Roorkee}\\
Roorkee, India \\
vishvendra@ece.iitr.ac.in}
}

\maketitle

\vspace{0em}
\begin{abstract}
Quantum dot-based spin qubits are resilient towards charge noise and are affected by magnetic noise only. However, environmental interaction leads to decoherence in these qubit systems. The external control parameters are directly related to the magnitude of decoherence. This in turn limits the range of values of those parameters for which operations can be done with high fidelity. In this work, using a model of quantum dot spin qubit system, we investigate the impact of varying ac magnetic fields on suppression of decoherence. We report an increment in the usable range of static magnetic field value using our technique.
\end{abstract}

\vspace{0em}
\section{Introduction}
A lot of work is being done in recent times to find a suitable system to physically implement qubits, which are the fundamental building blocks of quantum computation. Some of the leading candidates are superconducting qubits, trapped ion qubits, NV$^{-}$ center based qubits etc. One of the leading contenders among them is spin qubits based on planar quantum dot systems~\cite{loss1998quantum}. It is implemented using spins of electrons confined in the quantum dots. In a single qubit system, a static magnetic field is applied to a single confined electron. This causes a split in the degenerate spin energy level, thus giving rise to a two energy-level system. These two energy levels are used as two states of qubit. An ac magnetic field having a frequency equal to the transition frequency between the two energy levels is used to implement gate operations.
But like many other qubit systems, these spin qubits also suffer from noise due to decoherence. Interaction of the confined electrons with the environment is the major cause of decoherence in these systems and this in turn decreases the fidelity of the gate operations. The two major physical phenomena causing decoherence are hyperfine noise arising from nuclear spin fluctuations and phonon interaction mediated by spin-orbit interaction coupling ~\cite{taylor2006hyperfine,amasha2008electrical,khaetskii2001spin,san2006geometrical,marquardt2005spin,borhani2006spin}.

In this work, we investigate the effect of change of ac magnetic field value on the fidelity of single qubit $NOT$ gate operation in presence of decoherence. From this study, we show that the range of static magnetic field for which high-fidelity gate operation can be carried can be increased by manipulating the ac magnetic field. The paper is structured as follows: in section \MakeUppercase{\romannumeral 2}, we  discuss the model used to simulate the system dynamics including decoherence. In section \MakeUppercase{\romannumeral 3}, we discuss the results followed by conclusion.

\section{Simulation Methodology}

This section provides the methodology followed to simulate the effect of ac field value on fidelity of gate operations for a single quantum dot based spin qubit system in presence of decohrence. We use the Lindblad Master Equation to model the dynamics of our system, that is given as follows:
\begin{equation}\label{Master_Eqaution}
 \frac{d\rho(t)}{dt}= L_0 + L_D.
\end{equation}

\begin{equation}\label{Master_Equation_Cohrent_Part}
 L_0= -\frac{i[H(t), \rho(t)]}{\hbar}.
\end{equation}

The terms $L_0$ and $L_D$ capture the coherent and non-coherent parts of the dynamics of the quantum system, respectively. $L_0$, in general, is given by Eq.~\ref{Master_Equation_Cohrent_Part} where $\rho(t)$ denotes the state of the quantum system at an arbitrary time instant $t$ and $H(t)$ is the Hamiltonian for the closed system. The Hamiltonian $H(t)$ has the general form $H = -g \overrightarrow \mu . \overrightarrow B$. For our system, this expression is given by Eq.~\ref{Hamiltonian}
\begin{equation}\label{Hamiltonian}
 H(t) = -g\mu B_{static}\sigma_z - g\mu B_{ac}(cos(\omega t )\sigma_x - sin(\omega t )\sigma_y)
\end{equation}

where $g$ is the gyro-magnetic ratio, $\mu$ is the magnetic moment of electron and $\sigma_i$'s are the Pauli matrices where $i=\{x,y,z\}$. The non-coherent part of the Master Equation $L_D$ has the form given below:
\begin{equation}\label{Non_cohrent_master}
 L_{D}=\sum_{n}\frac{1}{2}( 2C_{n}\rho(t)C_{n}^\dagger-\rho(t)C_{n}^\dagger C_{n}-C_{n}^\dagger C_{n}\rho(t))
\end{equation}

where the operator $C_n$ corresponds to the $n^{th}$ decoherence operator due to hyperfine noise and phonon interaction.

\section{Results}

Following the simulation methodology discussed in the earlier section we obtain our results. First, we initialize our system to $\ket{\downarrow}$ state and then apply a $\pi$ pulse. After that we measure the probability of finding the qubit in $\ket{\uparrow}$ state. We do this for different values of the static magnetic field keeping the ac magnetic field constant.
This whole sequence is repeated for different values of ac magnetic field.
\vspace{0em}
\begin{figure}[htbp]
\centering
\includegraphics[scale=0.06]{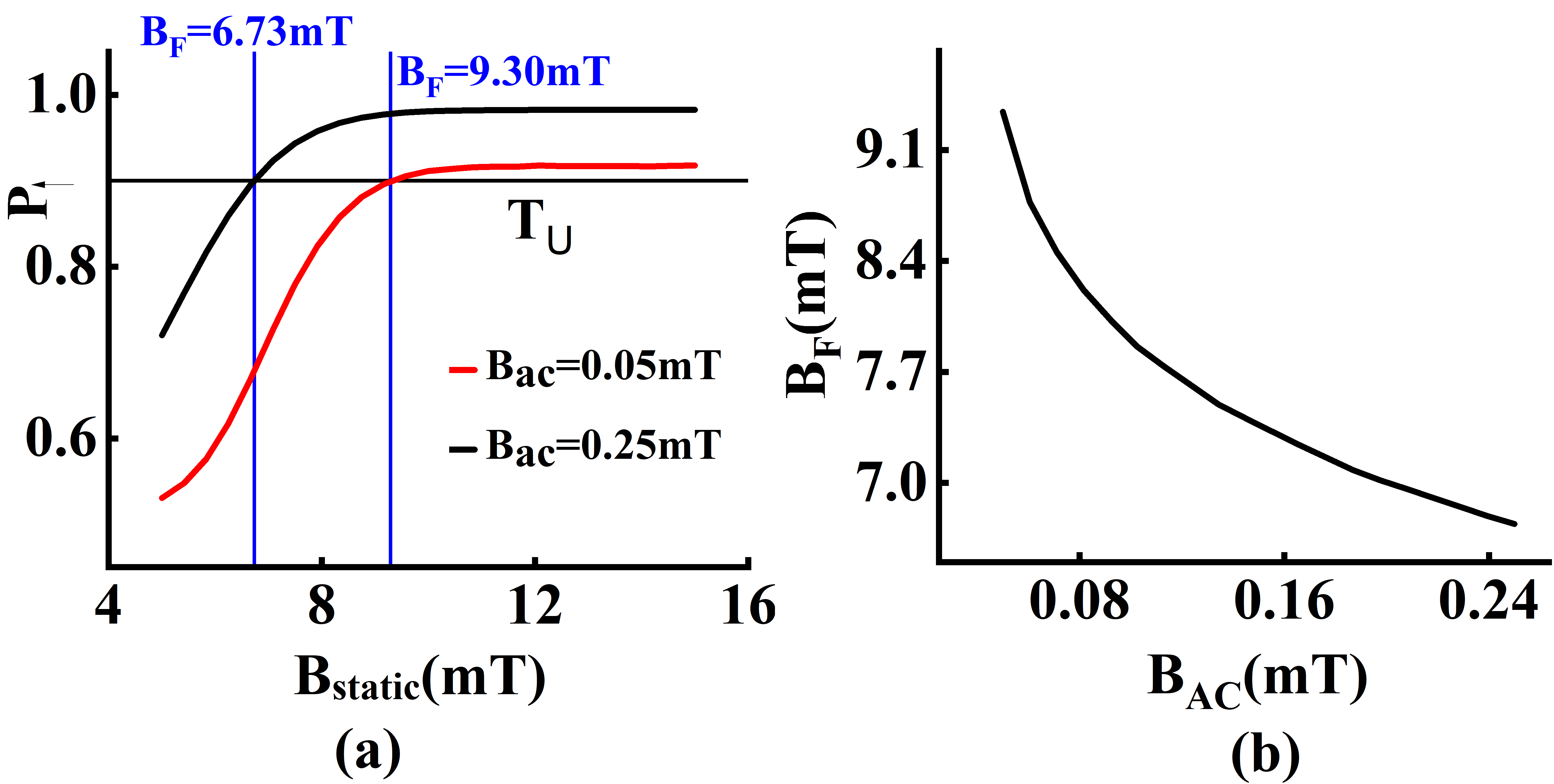}
\vspace{0em}
\caption{(a) Probability $(P_\uparrow)$ of finding the spin in $\ket{\uparrow}$ state after application of the $\pi$-pulse as function of varying static magnetic field $(B_{static})$. (b) Static magnetic field value for which probability $P_\uparrow$ drops to $0.9$ as function of ac magnetic field value. In this regime of $B_{static}$, hyperfine noise is the dominant decoherence mechanism.}
\label{results_for_low_B} 
\end{figure}

For the results obtained in Fig.~\ref{results_for_low_B}, we vary the static magnetic field $(B_{static})$ between $5$ mT to $15$ mT  keeping the ac field $(B_{ac})$ constant. This is repeated for various values of $(B_{ac})$ in the range of $0.05$ mT to $0.25$ mT. Decoherence due to hyperfine interaction is dominant for such low values of the static magnetic field~\cite{mehl2013noise}. It is observed that for a given value of the ac field, the lesser the value of $B_{static}$ more the decoherence as can be seen in Fig.~\ref{results_for_low_B}(a). The horizontal line $(T_{U})$ in Fig.~\ref{results_for_low_B}(a) denote the probability $P_\uparrow = 0.9$. The points represented by $B_F$ is the value of the static magnetic field for which the probability $(P_\uparrow)$ drops to $0.9$ for a certain value of $B_{ac}$. We can clearly see in Fig.~\ref{results_for_low_B}(a) that for higher values of $B_{ac}$, the values of $B_{F}$ is lower. In Fig.~\ref{results_for_low_B}(b), we plot the values of $B_F$ for different $B_{ac}$ values. It can be clearly seen from this figure that a higher value of $B_{ac}$, results in a lower the value of $(B_{F})$. This clearly shows that higher values of $B_{ac}$ can suppress the effect of decoherence due to hyperfine interaction to a certain extent.

\begin{figure}[htbp]
\centering
\includegraphics[scale=0.06]{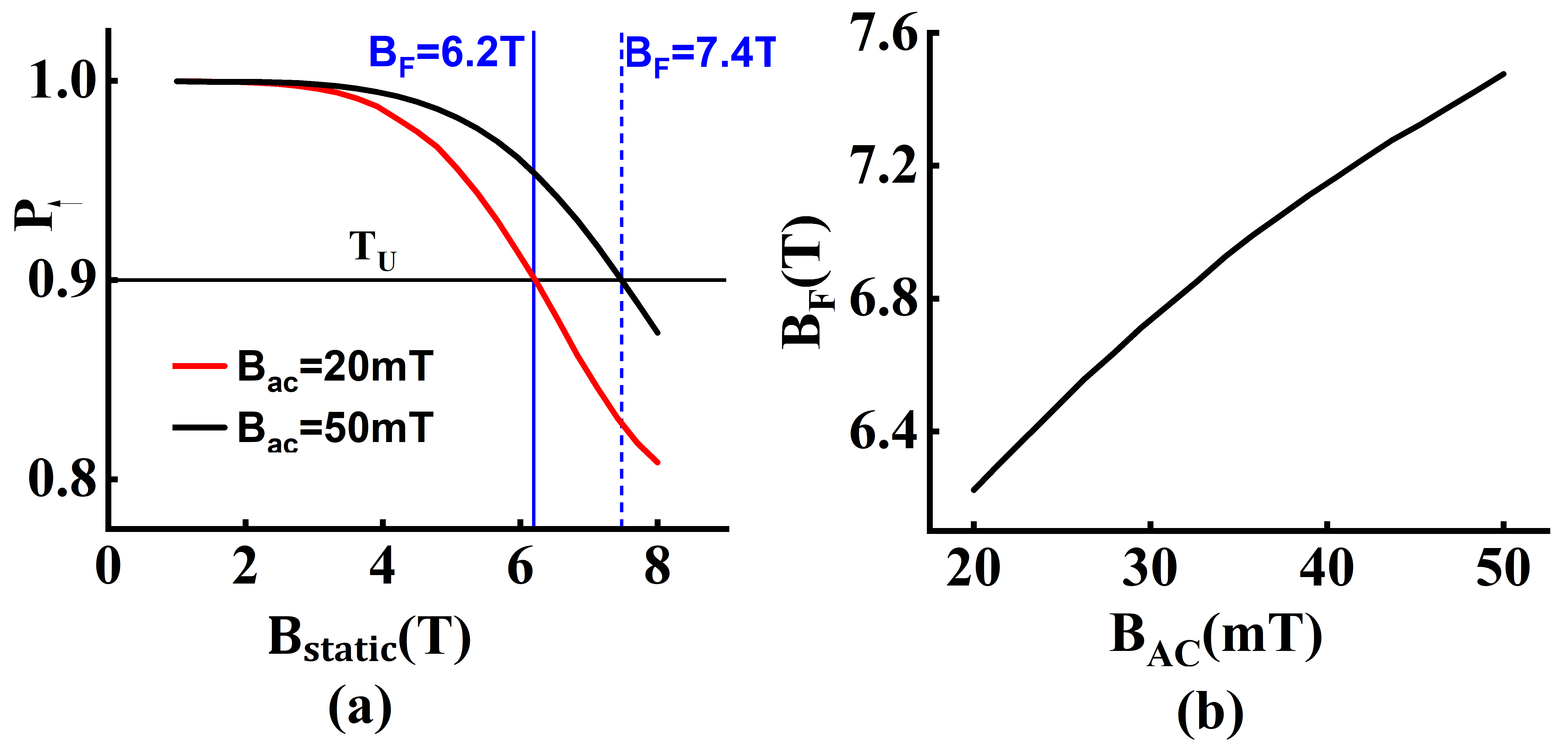}
\vspace{-1em}
\caption{(a) Probability $(P_\uparrow)$ of finding the spin in $\ket{\uparrow}$ state after application of the $\pi$-pulse as function of varying static magnetic field at high values, where phonon interaction is the dominant decoherence mechanism. (b) Static magnetic field value for which probability drops to $0.9$ of finding the spin in state $\ket{\uparrow}$ $(P_\uparrow)$ as function of ac magnetic field value where phonon interaction is the dominant decoherence mechanism.}
\vspace{0em}
\label{results_for_high_B} 
\end{figure}

For the results obtained in Fig.~\ref{results_for_high_B}, we vary the static magnetic field between $1$ T to $8$ T and the ac magnetic field is varied between $20$ mT and $50$ mT. Phonon interaction is dominant for such high values of the static magnetic field causing decoherence~\cite{mehl2013noise}. For a given value of the ac field, the higher the value of the static field $(B_{static})$, the more the decoherence as can be seen in Fig.~\ref{results_for_high_B}(a). The horizontal line and the points denoted as $B_F$ in Fig.~\ref{results_for_high_B}(a) denote the same quantities as earlier. We can see in Fig.~\ref{results_for_high_B}(a) that for higher values of ac field $(B_{ac})$ the values of $B_F$ is higher. In Fig.~\ref{results_for_high_B}(b), we plot the values of $B_F$ for varying $B_{ac}$ values. It can be clearly seen from this figure that for higher values of $B_{ac}$ we can work with higher values of $B_{static}$. So, it can be said that higher values of $B_{ac}$ can suppress the effect of decoherence due to phonon interaction to a certain extent.

In summary, we observe that a higher value of ac field mitigate decoherence at both high and low magnetic fields. We think this might be due to the fact that raising the value of the ac magnetic field decreases the $\pi$-pulse time (faster gate operation). Therefore, our system gets much lesser time to interact with its environment and this leads to lesser decoherence and a better performance of the system.

\section*{Conclusion}
 We have shown that we can obtain a high fidelity $NOT$ gate for much larger range of static magnetic field by increasing the ac magnetic field. This allows us to have a greater flexibility with both control parameters. Furthermore, we plan to undertake a similar study for multi-spin systems and other gates as well.



\bibliographystyle{IEEEtran}
\bibliography{IEEEabrv,Bibliography}

\end{document}